\documentclass{emulateapj}

\newcommand{\cxlong}{CXOU J013651.1$+$154547}
\newcommand{\cx}{M74 X-1}
\newcommand{\grs}{GRS 1915$+$105}

\begin{document}

\title{M74 X-1 (CXOU J013651.1+154547): AN EXTREMELY VARIABLE ULTRALUMINOUS
X-RAY SOURCE}

\author{M.I. Krauss\altaffilmark{1,3}, R.E. Kilgard\altaffilmark{1,2},
M.R. Garcia\altaffilmark{1}, T.P. Roberts\altaffilmark{2}, and 
A.H. Prestwich\altaffilmark{1}}
\altaffiltext{1}{Harvard-Smithsonian Center for Astrophysics}
\altaffiltext{2}{University of Leicester}
\altaffiltext{3}{Massachusetts Institute of Technology}

\begin{abstract}
 
Ultraluminous X-ray sources (ULXs) have been described variously as the most luminous normal X-ray binaries, as hypernovae, and as ``intermediate-mass'' black holes with masses of hundreds to thousands of solar masses.  We present results on \cx\ (\cxlong), a ULX in the nearby spiral galaxy M74 (NGC 628), from observations by {\it Chandra} and {\it XMM-Newton}. \cx\ is a persistent source which exhibits extreme variability and changes in spectral state on timescales of several thousand seconds.  Its variability timescales and behavior resemble some Galactic microquasars.  This suggests that the emission mechanism may be related to relativistically beamed jets, and that \cx~could be an extragalactic ``microblazar''---a microquasar whose jet axis is aligned with our line of sight.  We also note that its spectrum is consistent with the presence of a low-temperature disk blackbody component, which, assuming it is due to radiation from an accretion disk, could indicate that \cx\ contains an intermediate-mass black hole. 

\end{abstract}


\section{Introduction}

Ultraluminous X-ray sources (ULXs) are historically defined as point sources whose luminosities exceed the Eddington luminosity for a 10 $\rm M_{\odot}$ compact object, $\rm L_x\ \simeq\ 10^{39}\ erg\ s^{-1}$. Although they were discovered in the {\it Einstein} era \citep{fabbiano89}, and more than 100 ULX candidates were subsequently catalogued by the {\it ROSAT} HRI \citep{roberts00,colbert02}, much about the nature of these sources was unknown until recent times.  Observatories such as {\it ASCA} provided valuable information about the spectra of ULXs \citep[see, e.g.,][]{makishima00}; however, except in the cases where the sources were significantly variable, there remained the possibility that what appeared to be a ULX was really a group of spatially unresolved sources. 

The ability to spatially separate ULXs from their surrounding environs
is now possible with the superb resolution of the {\it Chandra
X-ray Observatory}, and both {\it Chandra} and {\it XMM-Newton} can
perform high quality X-ray spectroscopy and temporal analysis.  This
affords us detailed insight into the nature of ULXs:  we now know,
for example, that probably 10\% or fewer ULXs are associated with
recent supernovae \citep{roberts02a}, and that the others are, in
fact, isolated sources.  Although some ULXs are found in elliptical
galaxies \citep{colbert02},  most reside in star forming and
starburst galaxies, implying that they are associated
with recent star formation (though the exact nature of this link is unclear).  

A number of models have been proposed to explain the high X-ray luminosity of ULXs, 
and it is likely that more than one is required to explain all types of
observed ULX activity.  
Some ULXs may be extragalactic microquasars
\citep{kording02,foschini03,georganopoulos02,king01a},  in which case the observed
high luminosities would be due to relativistic beaming.  Others
may be  stellar-mass black holes with super-Eddington radiation \citep{begelman02}.   
Some are almost certainly background AGN that happen to fall along the line-of-sight
to the host galaxy  \citep{irwin04}.   One of the most intriguing 
suggestions is that they are intermediate-mass black holes \citep[IMBHs, see
e.g.][]{colbert99,makishima00,miller03}.  Aside from their high
luminosities, the low-temperature blackbody component in some ULX
spectra can be used to argue for the  presence of an IMBH
\citep{miller03}.   In this scenario, the blackbody spectrum is assumed to arise from
an accretion disk, and its temperature can be used to infer the radius of the
disk's inner edge and thus a mass for the central object.

In this paper we focus on observations of a specific ULX, \cx\
(\cxlong).  The first X-ray imaging observation of M74, which is at a distance of 8.7 Mpc \citep{huchra99}, was obtained with the {\it Einstein} HRI on 1980 January 5.  Though it is not included
in the {\it Einstein} source catalog, there is a source present at the
location of \cx\ with a 0.3--10 keV flux of ${\rm 7.6 \times 10^{-14}\
erg\ cm^{-2}\ s^{-1}}$ (14 total counts), corresponding to a luminosity of ${\rm 7
\times 10^{38}\ erg\ s^{-1}}$.  M74 was also observed by {\it ASCA} on 1995 July 28, but no source is detected at the location of \cx.  We derive an upper limit to the 0.2--10 keV flux of $\rm 7 \times 10^{-14}\ erg\ cm^{-2}\ s^{-1}$ (corresponding to a luminosity of $6 \times 10^{38} \rm ~erg~cm^{-2}~s^{-1}$).  No pointed {\it ROSAT} observations were performed.

The {\it Chandra} observations of M74 were conducted as part of a Cycle 2 large project to study the point source population of nearby, face-on spiral galaxies \citep{prestwich01}.  \cx, one of the objects in a subsample of ultraluminous point sources, was discovered to be extremely variable
and exhibit strong flares with a characteristic timescale of several thousand seconds.  In Section~\ref{xray}, we report on the recent X-ray observations of \cx; in particular, its spectral characteristics and variability.  In Section~\ref{opt_radio} we present optical and radio observations of M74.  In Section~\ref{disc}, we discuss \cx's location in M74, as well as its observed spectral and variability properties; and compare our findings with Galactic microquasars and with other ULXs.

\section{X-ray Observations and Data Analysis}
\label{xray}

{\it Chandra} has observed M74 twice, on 2001 June 19 and 2001 October 19.  Each observation utilized the ACIS back illuminated S3 chip, and basic processing was done by the {\it Chandra} X-ray Center (CXC).  There are two small background flares in the first observation during which background rates jump to two and three times the quiescent level.  These flares were not filtered out for analysis because they contribute a negligible ($\sim$~0.3) number of counts to the source data presented here.  The total duration of the ``good time intervals'' (GTIs) for the June dataset is 46.4 ks.  There are no background flares in the October observation, whose GTIs sum to 46.2 ks.  Both datasets have been filtered to include only energies in the range 0.3--10.0 keV. 

{\it XMM-Newton} observed M74 for 36 ks on 2002 February 2.  The data from MOS1, MOS2, and the PN were reduced individually, then fit simultaneously during spectral analysis, but were summed together for the temporal analysis. Data reduction was performed using SAS (Science Analysis Software) v5.3.  Several significant background flares were removed for analysis, resulting in the loss of $\sim 7$ ks of data.  The data were also filtered for ``good'' events plus pattern $\leq$ 12 (MOS) and $\leq$ 4 (PN), and for the energy range 0.2--10.0 keV.  The total duration of the GTIs are, for MOS1 and MOS2, 28.9 ks; and for the PN, 28.7 ks.  

\subsection{Spectral Fitting and Energetics}
\label{spec_fit}

Spectral fits were performed using {\it Sherpa}, the fitting tool included in the CIAO data analysis package.  Background data were subtracted prior to fitting.  {\tt powell} was used for optimization, and Chi Data Variance was the statistic function employed.  The fits to the {\it Chandra} ({\it XMM}) data were performed on data grouped in 15 (20) count bins.   The {\it Chandra} ARFs were corrected for the ACIS low energy QE degradation using the {\tt corrarf} program.\footnote{Information about {\tt corrarf} can be found at http://cxc.harvard.edu/cal/Links/Acis/acis/Cal\_prods/qeDeg/.}  In the {\it XMM} observation, \cx\ is 6$\arcmin$ off-axis and close to chip gaps on all the detectors, so the tools {\tt rmfgen} and {\tt arfgen} were used to produce correct response files for fitting. 

For the 2001 June 19 ({\it Chandra}) observation, there are a total of 347 source counts, of which approximately seven are due to background contamination. The 2001 October 19 ({\it Chandra}) observation contains a total of 963 source counts, of which approximately eight are due to background.  Both source ellipses are of a size that is expected to contain 98\% of the source energy, given their off-axis angle (3\farcm6 for June, and 3\farcm8 for October).  For the 2002 February 2 ({\it XMM-Newton}) observation, a source-free background region on the same chip as the source was used for background determination.  After background subtraction, there are a total of $\sim 650$ counts in the PN spectrum and $\sim 430$ counts in the MOS spectra.  

\begin{figure}
\resizebox{!}{3.3in}{\includegraphics{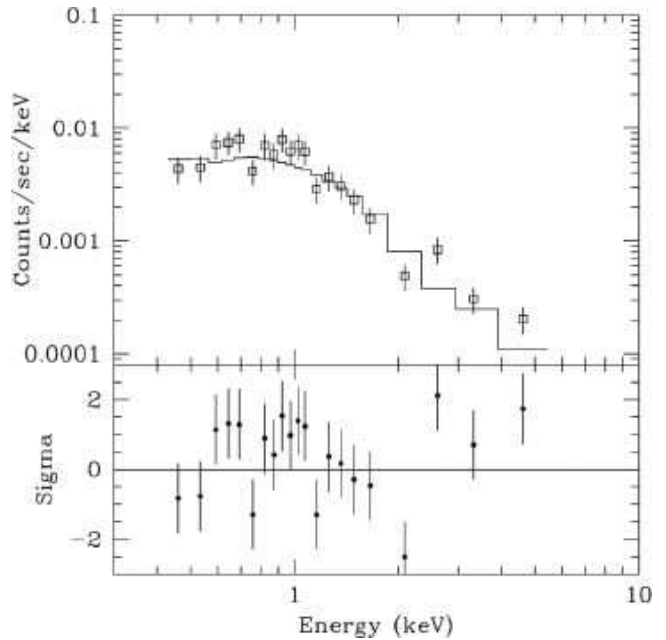}}
\caption{2001 June 19 {\it Chandra} spectrum with the absorbed (Galactic $N_{\rm H}$) power-law fit overplotted and residuals shown below.
\label{bestfit_june}}
\end{figure}

At its peak brightness in the {\it Chandra} observations, the 0.3--10.0 keV count rate is 0.11 cts s$^{-1}$, corresponding to a source flux of ${\rm 1.3\times10^{-12}~erg~cm^{-2}~s^{-1}}$, or a luminosity of ${\rm 1.2\times10^{40}~erg~s^{-1}}$.  To verify that the spectrum is not affected by pileup, we used the CIAO tool {\tt mkpsf} to create a PSF image at the location of \cx.  The inner 3 $\times$ 3 pixel region of the PSF contains 43.4\% of the total flux, whereas this would be 88.6\% were the source at the aimpoint of the detector. Using the pileup algorithm implemented in {\tt PIMMS} (Portable, Interactive Multi-Mission Simulator), the source at its brightest (0.11 cts s$^{-1}$) is 3.2\% piled, while on average (a rate of 0.04 cts s$^{-1}$) it is 1.2\% piled.  This will not cause a statistically significant change in the shape of the spectrum.

\begin{figure}
\resizebox{!}{3.3in}{\includegraphics{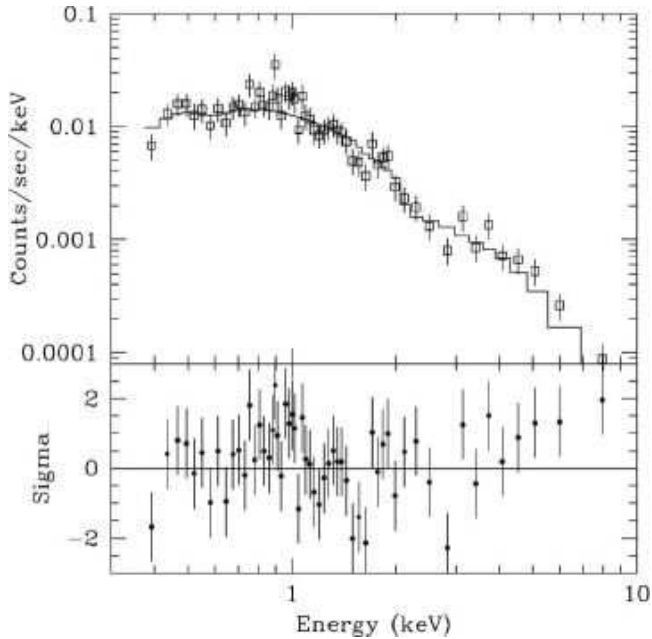}}
\caption{2001 October 19 {\it Chandra} spectrum with the absorbed (Galactic $N_{\rm H}$) power-law fit overplotted and residuals shown below.
\label{bestfit_oct}}
\end{figure}

\begin{deluxetable*}{cccccccccccccc}
\tabletypesize{\scriptsize}
\tablewidth{0pt} 
\tablecaption{Power-law and Power-law Plus Disk Blackbody Spectral Fits\tablenotemark{a}}
\tablehead{ 
\colhead{Observation}  & & \colhead{$N_{\rm H}$} & & \multicolumn{3}{c}{Power-law}  & & \multicolumn{3}{c}{Disk Blackbody}  & &
 \colhead{$\chi^{2}_{\nu}$ (DOF, sig.)} \\
 \cline{1-1} \cline{3-3} \cline{5-7} \cline{9-11} \cline{13-13}\\
 &  & & & \colhead{$\Gamma$}  &  \colhead{Amplitude\tablenotemark{b}}
 & \colhead{Flux\tablenotemark{c}} & & \colhead{kT$_{in}$}  &   \colhead{Norm}  & \colhead{Flux\tablenotemark{c}} & & \\
   & & \colhead{({\tiny $\times 10^{20}$})} & & & \colhead{({\tiny
$\times 10^{-5}$})} & \colhead{({\tiny $\times 10^{-13}$})} &
\colhead{} & \colhead{(keV)} & & \colhead{({\tiny $\times 10^{-13}$})}
& & \colhead{}
}
\startdata 
2001 June 19 & & 4.8 & & $2.25_{-0.18} ^{+0.19}$ & $0.96\pm0.09$  & 0.49 & & \nodata & \nodata & \nodata & & 1.67 (19, N/A) \\
 & & $4_{-4}^{+13}$ & & $2.19_{-0.38}^{+0.76}$ & $0.92_{-0.20}^{+0.52}$ & 0.48 & & \nodata & \nodata & \nodata & & 1.75 (18, $0.2\sigma$\tablenotemark{\S}~) \\
 & & 4.8 & & $0.6_{-1.3}^{+0.9}$ & $0.17_{-0.14}^{+0.31}$ & 0.45 & & $0.29_{-0.06}^{+0.07}$ & $0.22_{-0.12}^{+0.31}$ & 0.25 & & 1.25 (17, $2.1\sigma$\tablenotemark{\S}~) \\
 & & $7_{-7}^{+16}$ & & $0.7_{-1.6}^{+0.9}$ & $0.19_{-0.17}^{+0.38}$ & 0.45 & & $0.27_{-0.10}^{+0.13}$ & $0.32_{-0.26}^{+5.6}$ & 0.27 & & 1.32 (16, $2.0\sigma$\tablenotemark{\dag}~) \\
\tableline
\tableline
2001 Oct 19: All & & 4.8 &  &$2.02\pm0.10$ & $2.72\pm0.15$ & 1.52 & & \nodata & \nodata & \nodata & & 1.23 (55, N/A) \\
 & & $2.0_{-2.0}^{+3.6}$ & & $1.88_{-0.17}^{+0.21}$ & $2.45_{-0.28}^{+0.38}$ & 1.49 & & \nodata & \nodata & \nodata & & 1.22 (54, 1.1$\sigma$\tablenotemark{\S}~) \\
 & & 4.8 & & $1.1_{-0.5}^{+0.4}$ & $1.0_{-0.5}^{+0.6}$ & 1.35 & & $0.27\pm0.05$ & $0.6_{-0.3}^{+0.6}$ & 0.54 & & 0.96 (53, 3.5$\sigma$\tablenotemark{\S}~) \\  
 & & $6_{-6}^{+12}$ & & $1.15_{-0.71}^{+0.48}$ & $1.08_{-0.70}^{+0.87}$ & 1.36 & & $0.26_{-0.10}^{+0.13}$ & $0.8_{-0.6}^{+12.8}$ & 0.56 & & 0.98 (52, 3.3$\sigma$\tablenotemark{\dag}~) \\
 \tableline
Bright: & & 4.8 & & $1.77_{-0.16}^{+0.17}$ & $5.18_{-0.48}^{+0.47}$ & 3.39 & & \nodata & \nodata & \nodata & & 1.70 (27, N/A) \\
 & & $0^{+2.8}_{-N/A}$ & & $1.58^{+0.17}_{-0.14}$ & $4.44_{-0.40}^{+0.54}$ & 3.44 & & \nodata & \nodata & \nodata & & 1.58 (26, $1.7\sigma$\tablenotemark{\S}~) \\
 & & 4.8 & & $0.67_{-0.61}^{+0.50}$ & $1.6_{-0.9}^{+1.4}$ & 3.97 & & $0.29_{-0.07}^{+0.08}$ & $1.0_{-0.6}^{+1.8}$ & 1.09 & & 1.16 (25, $2.9\sigma$\tablenotemark{\S}~) \\
 & & $13_{-13}^{+19}$ & & $0.85_{-0.88}^{+0.54}$ & $2.0_{-1.5}^{+2.0}$ & 3.93 & & $0.22_{-0.08}^{+0.18}$ & $4.9_{-4.7}^{+390}$ & 1.7 & & 1.19 (24, $2.5\sigma$\tablenotemark{\dag}~) \\
 & & 4.8 & & $0.83_{-0.58}^{+0.55}$\tablenotemark{\ddag} & $1.8_{-1.0}^{+1.6}$ & 3.65 & & $0.31_{-0.05}^{+0.05}$\tablenotemark{\ddag} & $0.69_{-0.30}^{+0.58}$ & 1.01 & & 1.20 (25) \\
\tableline
Faint:\tablenotemark{d}& & 4.8 & & $2.29_{-0.14}^{+0.15}$ & $1.83_{-0.14}^{+0.14}$ & 0.93 & & \nodata & \nodata & \nodata & & 0.86 (26, N/A) \\
 & & $7.1_{-7.1}^{+8.9}$ & & $2.40_{-0.40}^{+0.49}$ & $1.99_{-0.51}^{+0.72}$ & 0.99 & & \nodata & \nodata & \nodata & & 0.88 (25, $0.6\sigma$\tablenotemark{\S}~) \\
 & & 4.8 & & $0.83_{-0.58}^{+0.55}$\tablenotemark{\ddag} & $0.29_{-0.17}^{+0.37}$ & 0.58 & & $0.31_{-0.05}^{+0.05}$\tablenotemark{\ddag} & $0.33_{-0.16}^{+0.34}$ & 0.48 & & 1.13 (26) \\
\tableline
\tableline
2002 Feb 2 & & 4.8 & & $1.96_{-0.09}^{+0.10}$ & $2.38\pm0.30$ & 1.4 & & \nodata & \nodata & \nodata & & 1.09 (59, N/A) \\
 & & $1.1\pm2.7$ & & $1.77\pm0.17$ & $2.09\pm0.34$ & 1.4 & & \nodata & \nodata & \nodata & & 1.04 (58, $1.9\sigma$\tablenotemark{\S}~) \\
 & & 4.8 & & $1.35\pm0.39$ & $1.3\pm0.7$ & 1.3 & & $0.25\pm0.06$ & $0.6\pm0.6$ & 0.3 & & 0.92 (55, $2.6\sigma$\tablenotemark{\S}~) \\
 & & $1.2\pm6.3$ & & $1.24\pm0.52$ & $1.1\pm0.8$ & 1.3 & & $0.31\pm0.16$ & $0.2\pm0.4$ & 0.3 & & 0.92 (54, $2.1\sigma$\tablenotemark{\dag}~) \\
\enddata 
\label{fit_param}
\tablenotetext{a}{Upper and lower limits are the 90\% confidence range.  $N_{\rm H}$ was frozen to the Galactic value of $4.8\times10^{20}$ cm$^{-2}$ for fits where there is no $N_{\rm H}$ confidence range given; otherwise, $N_{\rm H}$ was allowed to vary without bound.}
\tablenotetext{b}{The amplitudes of the power-laws are the flux at 1.0 keV in units of photons keV$^{-1}$ cm$^{-2}$ s$^{-1}$.}
\tablenotetext{c}{0.3-10.0 keV flux in units of erg/cm$^{2}$/s.}
\tablenotetext{d}{Adding a disk blackbody component actually increased the value of $\chi^{2}_{\nu}$ for these data.}
\tablenotetext{\dag}{Significance is determined using an $F$-test comparing this fit to a power-law fit of the same data with $N_{\rm H}$ allowed to vary.}
\tablenotetext{\S}{Significance is determined using an $F$-test comparing this fit to a power-law fit of the same data with $N_{\rm H}$ frozen to the Galactic value.}
\tablenotetext{\ddag}{For these fits, the ``faint'' and ``bright'' data were fit using a common power-law slope and blackbody temperature, allowing only their normalizations to vary.}
\end{deluxetable*} 

\begin{figure}
\resizebox{!}{3.3in}{\includegraphics{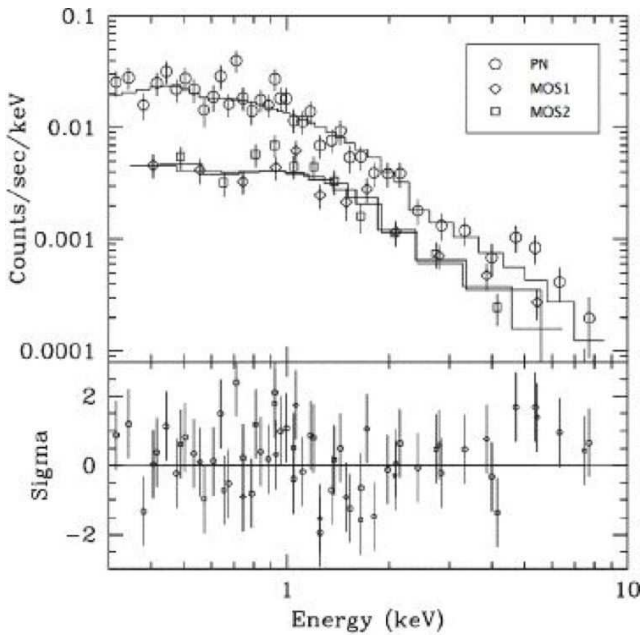}}
\caption{2002 Feb 2 {\it XMM-Newton} spectrum with the power-law model (Galactic $N_{\rm H}$) overplotted and residuals shown below.  PN data are plotted with circles, MOS1 data are plotted without a symbol, and MOS2 data are plotted with squares.
           \label{bestfit_feb}}
\end{figure}

\begin{figure}
\resizebox{!}{3.3in}{\includegraphics{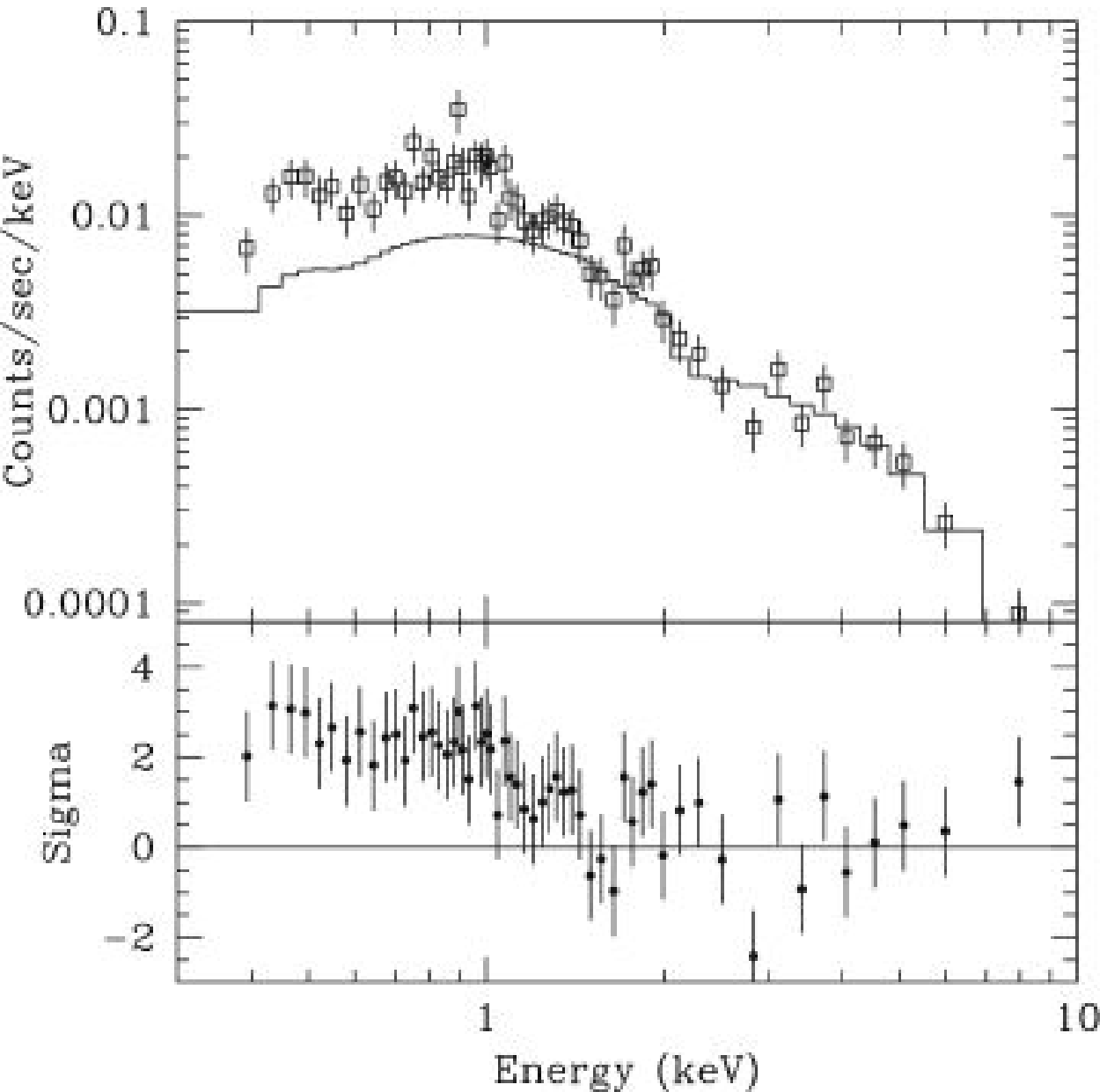}}
\caption{2001 October 19 {\it Chandra} spectrum with a power-law fit over the restricted energy range 1.5--10 keV overplotted and residuals shown below.  
\label{oct_excess}}
\end{figure}

\begin{figure}
\resizebox{!}{3.3in}{\includegraphics{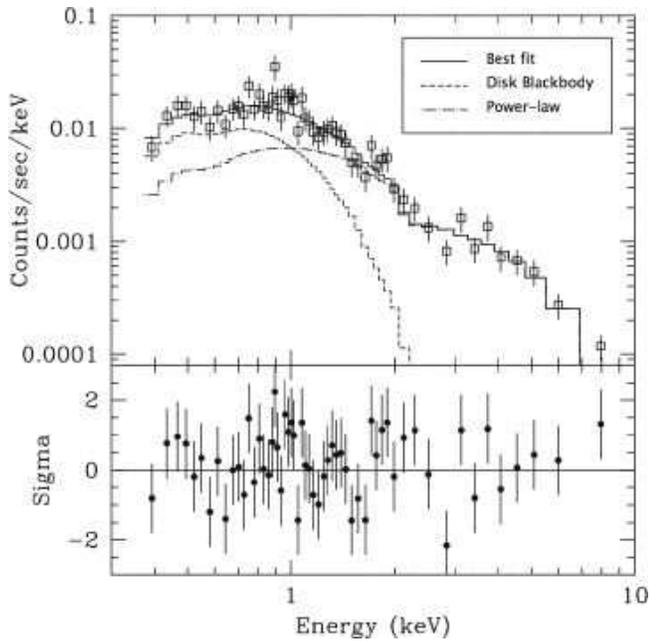}}
\caption{2001 October 19 {\it Chandra} spectrum fit by a power-law plus disk blackbody.  The fit and individual componenets are overplotted and residuals are shown below.  
\label{oct_excess_fit}}
\end{figure}

For initial spectral fitting, we tried a range of standard simple models, such as a disk blackbody, power-law, thermal bremsstrahlung, and MEKAL.  Of these models, only the power-law provides acceptable fits (see Figures~\ref{bestfit_june}, \ref{bestfit_oct}, and~\ref{bestfit_feb}).  However, the 2001 October 19 spectrum still has significant residuals beyond the power-law fit.  If we restrict the fitting range of the power-law to cover 1.5--10.0 keV, these residuals are manifested as a soft excess (see Figure~\ref{oct_excess}).  We were motivated by the spectra seen in Galactic X-ray binaries \citep[and in other ULXs, c.f.][]{miller03} to add a disk blackbody (DBB) component to the power-law, and using an $F$-test, we find that this improves the fit at the $3.3\sigma$ level (see Figure~\ref{oct_excess_fit}).  We find that the soft excess is similarly well-fit by a MEKAL model with low abundances and by a single-temperature blackbody spectrum.  Using an $F$-test to compare these models with the power-law fit, we determine that the addition of a MEKAL component is favored at the $3.7\sigma$ level, whereas the blackbody component is favored at the $3.1\sigma$ level.  We also fit the spectrum with a broken power-law model, but find that the parameters are not well constrained.  We present the fit parameters for all the datasets in Tables~\ref{fit_param} (power-law and power-law plus disk blackbody fits) and~\ref{fit_param_mk} (power-law plus MEKAL fits).

\begin{deluxetable*}{ccccccccccccccc}
\tabletypesize{\scriptsize}
\tablewidth{0pt} 
\tablecaption{Power-law Plus MEKAL Spectral Fits\tablenotemark{a}}
\tablehead{ 
\colhead{Observation}  & & \colhead{$N_{\rm H}$} & & \multicolumn{3}{c}{Power-law}  & & \multicolumn{4}{c}{MEKAL}  & &
 \colhead{$\chi^{2}_{\nu}$ (DOF, sig.)} \\
 \cline{1-1} \cline{3-3} \cline{5-7} \cline{9-12} \cline{14-14}\\
 &  & & & \colhead{$\Gamma$}  &  \colhead{Amplitude\tablenotemark{b}}
 & \colhead{Flux\tablenotemark{c}} & & \colhead{Plasma Temp.} & \colhead{Abund.} & \colhead{Norm}  & \colhead{Flux\tablenotemark{c}} & & \\
   & & \colhead{({\tiny $\times 10^{20}$})} & & & \colhead{({\tiny $\times 10^{-5}$})} & \colhead{({\tiny $\times 10^{-13}$})} & \colhead{} & \colhead{(keV)} & & \colhead{({\tiny $\times 10^{-6}$})} & \colhead{({\tiny $\times 10^{-13}$})} & & \colhead{}}
\startdata 
2001 June 19 & & 4.8 & & $1.93_{-0.35}^{+0.33}$ & $0.76\pm0.18$ & 0.45 & & $0.25_{-0.06}^{+1.1}$ & 1 & $3.0_{-2.4}^{+2.3}$ & 0.06 & & 1.60 (17, $1.1\sigma$\tablenotemark{\S}~) \\
\tableline
2001 Oct 19: All & & 4.8 & & $1.93_{-0.14}^{+0.13}$ & $2.42_{-0.25}^{+0.24}$ & 1.43 & & $0.7_{-0.4}^{+0.2}$ & 1 & $2.8_{-1.7}^{+1.8}$ & 0.08 & & 1.15 (53, $1.9\sigma$\tablenotemark{\S}~) \\
 & & $0_{-N/A}^{+2.1}$ & & $1.68_{-0.11}^{+0.15}$ & $1.95_{-0.18}^{+0.26}$ & 1.38 & & $0.7_{-0.2}^{+0.2}$ & 1 & $3.3\pm1.5$ & 0.09 & & 1.03 (52, $2.9\sigma$\tablenotemark{\dag}~) \\
 & & 4.8 & & $0.7_{-0.6}^{+0.5}$ & $0.5_{-0.3}^{+0.6}$ & 1.30 & & $0.8\pm0.2$ & $0.02_{-0.01}^{+0.02}$ & $153_{-32}^{+28}$ & 0.72 & & 0.90 (52, $3.6\sigma$\tablenotemark{\S}~) \\
 & & $2.8_{-2.8}^{+5.1}$ & & $0.7\pm0.6$ & $0.6_{-0.4}^{+0.7}$ & 1.29 & & $0.8\pm0.2$ & $0.02_{-0.02}^{+0.04}$ & $126_{-60}^{+95}$ & 0.64 & & 0.90 (51, $3.7\sigma$\tablenotemark{\dag}~) \\
 \tableline
 Bright: & & 4.8 & & $1.40_{-0.29}^{+0.28}$ & $3.67_{-0.87}^{+0.89}$ & 3.44 & & $0.7\pm0.2$ & 1 & $11.7\pm5.7$ & 0.32 & & 1.40 (25, $2.1\sigma$\tablenotemark{\S}~) \\
  & & $0_{-N/A}^{+2.4}$ & & $1.30_{-0.25}^{+0.23}$ & $3.31\pm0.69$ & 3.47 & & $0.7\pm0.2$ & 1 & $9.9_{-4.7}^{+4.9}$ & 0.27 & & 1.24 (24, $2.3\sigma$\tablenotemark{\dag}~) \\
  & & 4.8 & & $0.44_{-0.63}^{+0.51}$ & $1.1_{-0.7}^{+1.1}$ & 4.02 & & $0.8\pm0.2$ & $0.04_{-0.03}^{+0.04}$ & $249_{-85}^{+77}$ & 1.29 & & 0.98 (24, $3.3\sigma$\tablenotemark{\S}~) \\
  & & $0_{-N/A}^{+8.1}$ & & $0.56_{-0.30}^{+0.48}$ & $1.3_{-0.8}^{+1.1}$ & 3.93 & & $0.8_{-0.2}^{+0.3}$ & $0.06_{-0.04}^{+0.06}$ & $160_{-70}^{+200}$ & 0.95 & & 1.00 (23, $3.0\sigma$\tablenotemark{\dag}~) \\
\enddata 
\label{fit_param_mk}
\tablenotetext{a}{Upper and lower limits are the 90\% confidence range.  $N_{\rm H}$ was frozen to the Galactic value of $4.8\times10^{20}$ cm$^{-2}$ for fits where there is no $N_{\rm H}$ confidence range given; otherwise, $N_{\rm H}$ was allowed to vary without bound.}
\tablenotetext{b}{The amplitudes of the power-laws are the flux at 1.0 keV in units of photons keV$^{-1}$ cm$^{-2}$ s$^{-1}$.}
\tablenotetext{\dag}{Significance is determined using an $F$-test comparing this fit to a power-law fit of the same data with $N_{\rm H}$ allowed to vary.}
\tablenotetext{\S}{Significance is determined using an $F$-test comparing this fit to a power-law fit of the same data with $N_{\rm H}$ frozen to the Galactic value.}
\end{deluxetable*}  

\begin{figure}
\resizebox{!}{3.3in}{\includegraphics{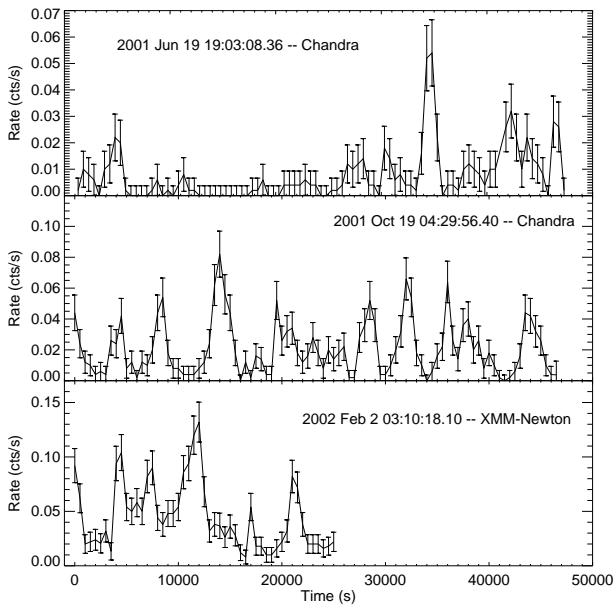}}
\caption{Lightcurves for all three observations.  The data are
           binned by 500 s, and the full energy range (0.3--10.0 keV) has
           been used.
\label{lightcurves}}
\end{figure}

\begin{figure}
\resizebox{!}{3.3in}{\includegraphics{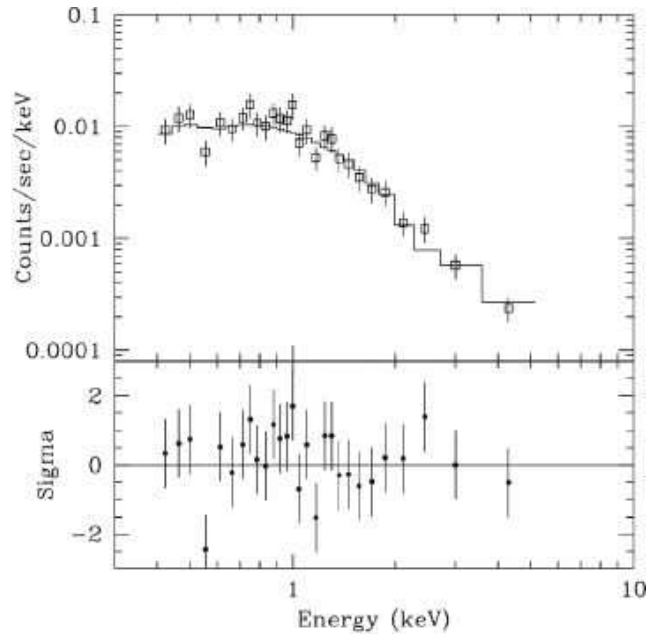}}
\caption{2001 October 19 ``faint'' {\it Chandra} spectrum with the power-law model 
           overplotted and residuals shown
           below.
\label{bestfit_faint}}
\end{figure}

\begin{figure}
\resizebox{!}{3.3in}{\includegraphics{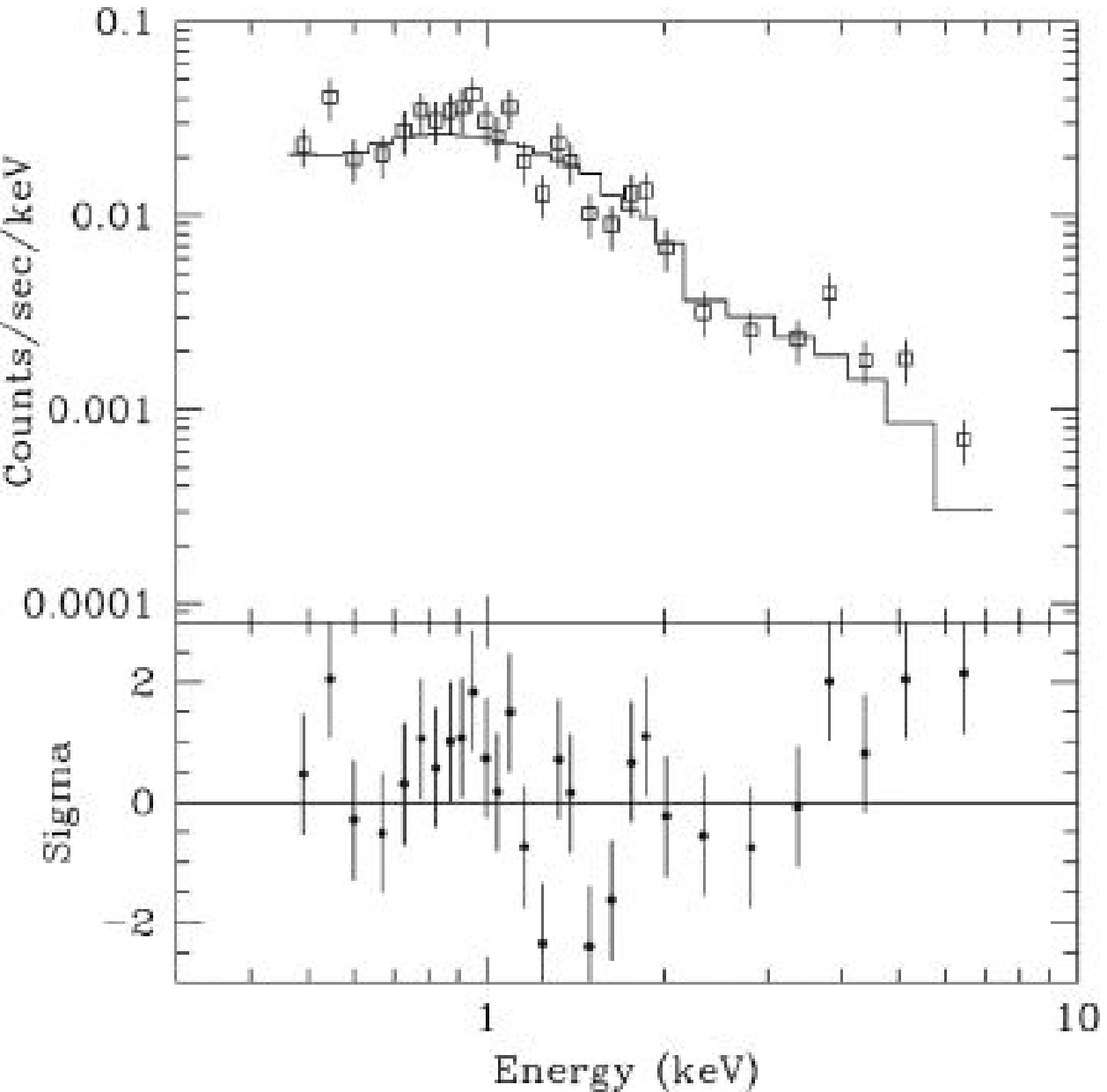}}
\caption{2001 October19 ``bright'' {\it Chandra} spectrum with the power-law model overplotted and residuals shown below.
\label{bestfit_bright}}
\end{figure}

To explore possible spectral variability in the temporally variable October data (the lightcurve is presented in Figure~\ref{lightcurves}), we separated it into ``bright'' (count rate $>$ 0.03 cts s$^{-1}$) and ``faint'' (count rate $<$ 0.03 cts s$^{-1}$) segments.  Both segments had comparable numbers of counts; 490 in the bright data and 469 in the faint data, allowing for similar statistical properties.  (There are an insufficient number of counts in the June observation to allow for a similar separation.)  Fitting with simple power-laws, we find that the bright data are somewhat harder than the faint data: the best-fit photon index for the bright data is $1.6_{-0.1}^{+0.2}$, whereas for the faint data it is $2.4_{-0.4}^{+0.5}$.  We also note that there appears to be a soft excess present in the bright data---as with the full dataset, this can be modeled variously with a disk blackbody, MEKAL component, and single-temperature blackbody.  Although the spectra appear to be different, it is possible that the physical parameters of the system (for example, the power-law slope and blackbody temperature) remain the same as the source brightens, and that only the contributions of different components vary.  To test this, we fit the bright and faint data simultaneously with an absorbed power-law plus a disk blackbody.  We used a common power-law slope and disk blackbody temperature, and allowed only the normalizations to vary.  Both datasets are consistent with a power-law slope of $\Gamma=0.8\pm0.6$ and a disk temperature of $0.31_{-0.05}^{+0.05}$ keV, but we must allow the amplitudes of both components to vary in order to get reasonable fits for the bright data.  Plots of the ``faint'' and ``bright'' spectra are presented in Figures~\ref{bestfit_faint} and~\ref{bestfit_bright}, and the results of spectral fitting are included in Tables~\ref{fit_param} and~\ref{fit_param_mk}.  

\begin{figure}
\resizebox{!}{3.3in}{\includegraphics{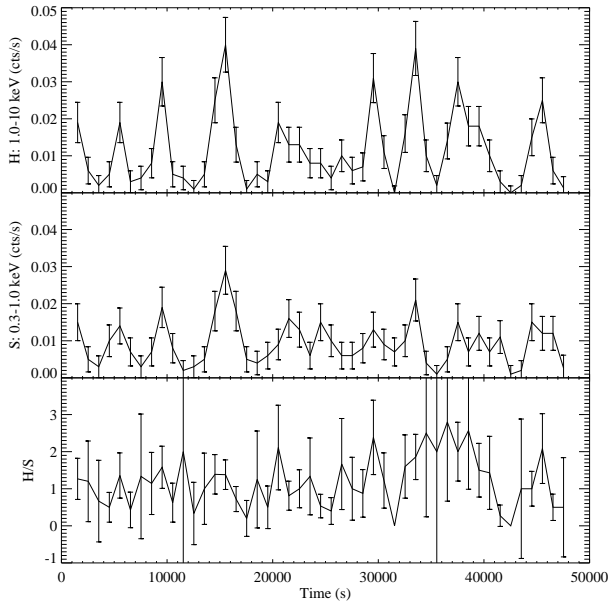}}
\caption{2001 October 19 lightcurves in hard (1.0--10.0 keV) and soft (0.3--1.0 keV) bands, with the hardness ratio plotted below.\label{energy_lightcurves}}
\end{figure}

\begin{figure}
\resizebox{!}{2.5in}{\includegraphics{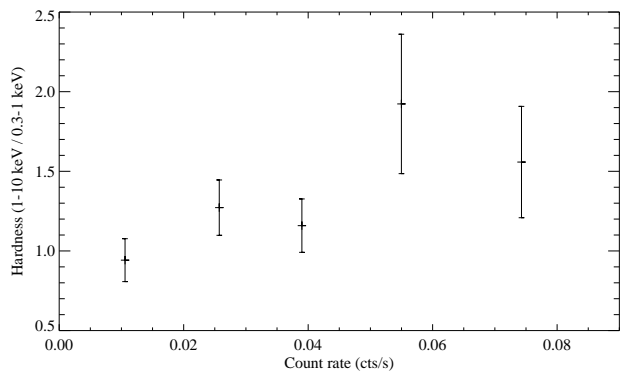}}
\caption{Hardness ratio (1.0--10.0 kev)/(0.3--1.0 keV) plotted against count rate for the 2001 October 19 data.  Counts have been summed over specific count rate intervals in order to improve the statistics of the hardness ratios.  A linear fit to the data shows that a line of slope 11 is preferred over a line of slope zero at the $1.7\sigma$ level.  
\label{hr_diagram}}
\end{figure}

To further explore the spectral variability, we plotted lightcurves of \cx\ in hard (1.0--10.0 keV) and soft (0.3--1.0 keV) bands.  These lightcurves, along with a plot of the hardness ratio, are presented in Figure~\ref{energy_lightcurves}.  Unfortunately, the limited number of counts (and therefore the statistical noise) hinders detailed interpretation.  Therefore, we also binned the data according to the source count rate, and calculated its hardness ratio.  There appears to be a trend toward increasing hardness with higher count rate, as can be seen in Figure~\ref{hr_diagram}, although a linear fit with non-zero slope is preferred over a flat distribution at only the $1.6\sigma$ level.

We performed a similar division of the 2002 February 2 dataset, but find that spectral hardening with increasing count rate is not present in these data. 

\subsection{Variability}

\cx\ is active and extremely variable in all three observations, with count rates sometimes increasing  by an order of magnitude on a timescale of several thousand seconds.  This variability is not caused or influenced by spacecraft dither or instrumental effects, and appears as a sequence of ``flares".  The rise and decay of these flares generally appear symmetric, with a few exceptions when a flare is followed by a period of ``noisy'' decay (for example, after the fourth and seventh flares in the 2001 October 19 data; see the lightcurves presented in Figure~\ref{lightcurves}). The source is relatively inactive during the first half of the 2001 June 19 observation, but exhibits several significant flares in the latter half.  It is active throughout the 2001 October 19 and 2002 February 2 observations; the former contains one particularly energetic flare during which the count rate rises from ($0.004 \pm 0.003$) cts s$^{-1}$ to ($0.08 \pm 0.01$) cts s$^{-1}$ in 2500 s.    

\begin{figure}
\resizebox{!}{2.6in}{\includegraphics{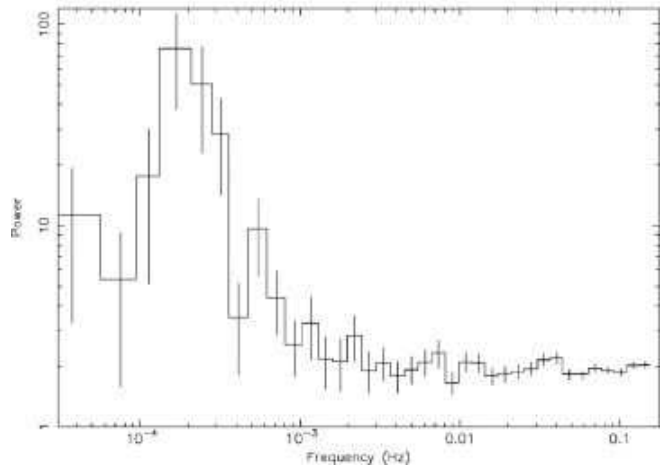}}
\caption{Normalized power density spectrum of the 2001 October 19 observation.  The FFT was performed with a bin time of 3.241 s, and the frequency bins were averaged to improve the statistics.
\label{fft_plot}}
\end{figure}

\begin{figure}
\resizebox{!}{3.3in}{\includegraphics{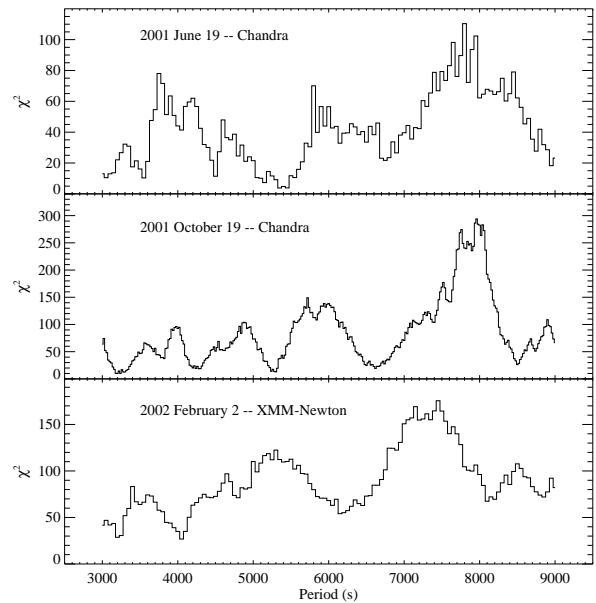}}
\caption{Period folding for all three observations.
           Folding was performed with the IRAF PROS routine {\tt
           period} from 3000-9000 s with 50 s binning.
\label{period_fold}}
\end{figure}

An FFT reveals that there are a number of significant extremely low-frequency peaks (which could be interpreted as quasi-periodic oscillations, or QPOs) in the power spectrum of the October observation (see Figure~\ref{fft_plot}).  However, there are no significant peaks at higher frequencies up to the Nyquist frequency of 0.156 Hz in either of the {\it Chandra} observations, both of which have frame times of 3.2 s.  Since the {\it XMM} PN data have time resolution of 73.4 ms (and a total of 660 counts), we searched it for higher frequency signals, but did not find anything significant.  Performing period folding using the IRAF task {\tt period} on all three of the observations produces $\chi^{2}$ values for a range of periods (see Figure~\ref{period_fold}).  There are broad peaks present in each, although there is not a single fundamental frequency.    

\section{Optical and Radio Data}
\label{opt_radio}

M74 was observed by the Gemini North telescope as part of its commissioning campaign with the GMOS detector on 2001 August 13--14.  Four images in each filter---$G$ (475 nm), $R$ (630 nm), and $I$ (780 nm)---were obtained and co-added for a total of 960, 480, and 480 s respectively.  Basic processing was done at the Gemini Observatory. To register the world coordinate system (WCS) of the $R$-band image, we first extracted bright stars using the {\tt SExtractor} program, then used the {\tt imwcs} routine (both from the WCSTools package\footnote{WCSTools documentation can be found at http://tdc-www.harvard.edu/software/wcstools/.}) in conjunction with the USNO-A2.0 catalog  to match 93 M74 sources to USNO-A2.0 objects.  Sources in the $G$- and $I$-bands were then centroided and these images were aligned to the $R$-band image using IRAF.  After this process was complete, the centroids of three bright X-ray sources in the field which have optical counterparts fell within $\pm~ 0\farcs2$ of them.  Within the $0\farcs2$ {\it Chandra} error circle there is no counterpart down to the limiting magnitude of $\rm m_V \simeq 25$ (calculated using the GMOS North Integration Time Calculator\footnote{This calculator can be found at http://www.gemini.edu/sciops/instruments/gmos/gmosIndex.html following the Performance and Use link.}, assuming a detection signal-to-noise threshold of 5.)  

We also obtained observations of M74 taken on 2001 December 16 with the Fred Lawrence Whipple Observatory (FLWO) 1.2 m telescope.  The galaxy was imaged in $UBVRI$ (Harris broadband filter set), SII, and H$\alpha$.  We note that there is no obvious ionization nebula present in either the SII or H$\alpha$ images, as is sometimes observed in the vicinity of ULXs \citep{pakull02,roberts03}.   A Very Large Array (VLA) observation at 1.425 GHz did not reveal any radio counterpart with an upper limit of 80 $\mu$Jy \citep{stockdale02}.  The X-ray flux density of the 2001 October 19 spectrum is $4\times 10^{-14} \rm~erg~cm^{-2}~s^{-1}~keV^{-1}$ at 1.0 keV, which gives a flux density of $1.7\times 10^{-2} \rm~\mu Jy$.  Given this X-ray flux, the lower limit to the X-ray--to--radio flux ratio is $\log[F_{\rm X}/F_{\rm R}] > -3.8$.

\begin{figure}
\resizebox{!}{3.9in}{\includegraphics{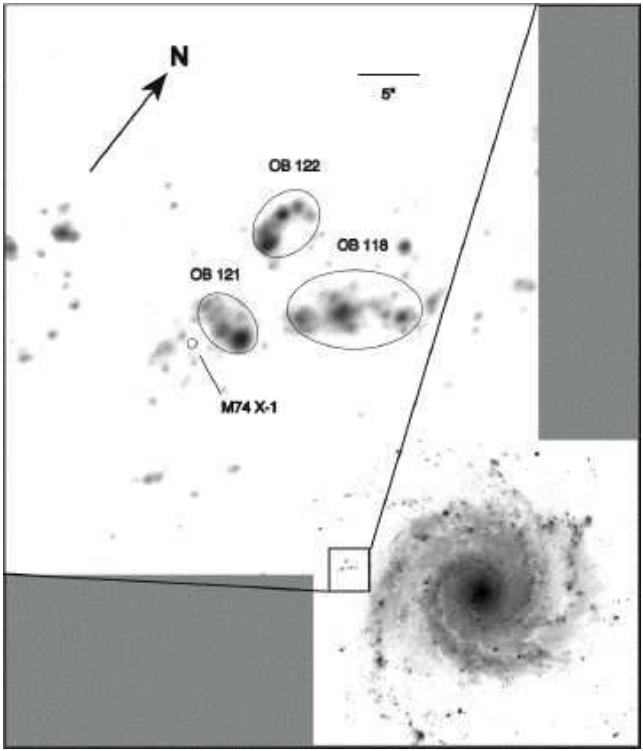}}
\caption{Gemini GMOS data (GRI image courtesy Gemini
           Observatory/GMOS team) overlaid with regions for OB associations 118, 121, and 122 \citep{ivanov92}
           and the 0.2$\arcsec$ positional uncertainty of \cx\ (small circle).
\label{source_optical}}
\end{figure}

\section{Discussion}
\label{disc}

\cx\ has some very unusual observational characteristics.  It is a ULX 
(peak luminosity of $1.2 \times 10^{40} \rm~erg~s^{-1}$) which is located near a young star cluster 
in M74.  Its spectrum is reasonably fit by a simple power law, but
in some observations there appears to be an excess at lower energies.  This excess can be fit by
a low temperature disk blackbody (0.3 keV) or a MEKAL model with low
abundances.  Its flux varies by an order of magnitude on timescales 
of half an hour, and there is tantalizing evidence for a QPO.  The
spectrum appears to harden as the source brightens, although the
significance of this effect is low.  In this section, we consider the physical interpretations
for these observations.

\subsection{Location of \cx}\label{location}

A major consideration in the study of ULXs is whether or not they actually reside in the galaxy with which they are coincident or whether they are, in fact, background or foreground sources.  If \cx\ were in the Milky Way, given its Galactic latitude of $-$45.7 and assuming it is 1.4 kpc above the Galactic plane (in the thick disk population), it would be at a distance of $\sim 2$ kpc.  If it were an LMXB, its counterpart would likely have 4 ${\rm \ge M_{V}\ge-4}$ \citep{vanparadijs94}, which at a distance of 2 kpc (and unobscured) would have 15.5 ${\rm \ge m_{V}\ge7.5}$.  Although the optical and X-ray observations were not simultaneous, the source seems to be X-ray active from June 2001 to February 2002, and two of the optical observations (Gemini and FLWO) were within this period.  No object brighter than the ${\rm m_{V}}=20.6$ OB association (or, in the Gemini data, any object at all within our error ellipse) is detected in any of the optical observations---we therefore consider it unlikely that \cx\ is a Galactic LMXB.  Furthermore, it is unlikely that \cx\ is a Galactic cataclysmic variable, since large-amplitude X-ray oscillations of this type are most often seen in polars and intermediate polars, and would then be related to the spin period of the white dwarf.  This would result in highly sinusoidal periodic variation.  However, the observed variability is only quasi-periodic, and in June, the flares are only observed toward the end of the observation.  Therefore, \cx\ most likely resides outside our Galaxy.  

It is also possible that \cx\ is a background active galactic nucleus
(AGN).  Using equations 1 and 2 of \citet{giacconi01}, and ignoring
extinction from M74, it can be seen that zero to two sources of this
flux would be expected on a single ACIS chip.  If it were an AGN, it
would most likely be a Narrow-Line Seyfert 1 (NLS1): these objects are
highly X-ray variable, and are often not radio-loud---but note that
they do have optical counterparts, unlike \cx.  Taking the detection
limit of $\rm m_V = 25$ as the magnitude of the optical counterpart,
the lower limit to the X-ray--to--optical flux ratio is $\log[F_{{\rm
X}}/F_{{\rm V}}]=2.3$ \citep{maccacaro82}.  Here, we assume no excess
absorption---which would increase this ratio---since the absorption
present in the spectra is always consistent with the Galactic value of
$N_{\rm H}$.  Among AGN, BL Lac objects have the highest X-ray to
optical flux ratio, ranging from $\log[F_{{\rm X}}/F_{{\rm V}}]=0.3$
to 1.7 \citep{stocke91}: this is significantly lower than what is
observed for \cx.  Furthermore, there are only a few AGN which
approach the variability exhibited by \cx; for example, IRAS
13244$-$3809 \citep{dewangan02,boller97} and PHL 1092
\citep{brandt99}.  Since \cx's extreme variability and lack of
detected radio or optical counterpart all argue against its
identification as a  background AGN, we conclude that it almost
certainly resides in M74. 

\subsection{Limits on the binary system from optical data}\label{limits}

Assuming that \cx\ is located in M74, the peak luminosity of
$\rm 1.2 \times 10^{40}~erg~s^{-1}$ implies that \cx\ most likely
contains a black hole, and probably has a binary companion.    At the 
distance of M74, the detection limit of $m_V \simeq 25$ corresponds to $\rm M_V 
\gtrsim -4.7$; thus any possible companion must be fainter than a B0III star \citep[$\rm M_V \sim -4.9$,][]{dejager87}.

The Gemini data with the {\it Chandra} positional error circle and nearby OB associations overlaid are shown in Figure~\ref{source_optical}.  \cx\ is near (but 
clearly not in) an OB association which has $\rm m_V = 20.6$ \citep[OB 121, see]
[]{ivanov92}, consistent with a group of approximately 25 O or B
stars.  This is interesting,  since ULXs are often located near star-forming regions \citep[see, e.g.][]{zezas02,goad02}.  Following a similar argument to that presented in \citet{zezas02}, we consider the possibility that \cx\ is a binary which was ejected from the nearby star cluster OB 121.  In this case, its projected separation of 2$\arcsec$ (corresponding to 85 pc) could be attained with a velocity in the plane of the sky of 30 km s$^{-1}$ for a period of approximately $3 \times 10^6$ years. This velocity is not unexpected for stellar-mass black-hole binaries, given the observed velocities of such binaries and radio pulsars within our Galaxy.  In fact, the velocity could be significantly higher; the Galactic microquasar GRO 1655$-$40 has a velocity of $\rm \sim 100\ km\ s^{-1}$ \citep{mirabel02}---adopting this value for \cx\ implies that it was ejected only $\sim 8 \times 10^{5}$ years ago.  The implied lifetime of $\sim 10^6$ years is also compatible with the lifetime of the X-ray emitting phase of a $M _{{\rm BH}} \sim10~M_{\sun}$  black-hole X-ray binary \citep{podsiadlowski03}.   Taking the average luminosity of the system to be ${\rm 10^{39}~erg~s^{-1}}$, the average $\dot M$ is given by the equation $\dot M_{{\rm ave}} = L_{{\rm ave}}/\eta c^{2}$.  Since $\eta \sim0.1$ for black hole systems, $\dot M \sim 2 \times 10^{-7}~M_{\sun}$ yr$^{-1}$.  Assuming that the primary is a black hole of $10~M_{\sun}$, this $\dot M$ implies that $M_{2} \gtrsim 8\ M_{\sun}$, and the optical data imply that $M_{2} \lesssim 20\ M_{\sun}$ \citep[the lower limit is taken from Figure 4 of][and the upper limit is approximately the mass of an 09V or B0III star, both of which are at the luminosity limit for detection in the GMOS data]{podsiadlowski03}.  However, we caution that the peak luminosity of \cx\ is an order of magnitude higher than the Eddington limit for a $10~M_{\sun}$ black hole; this is discussed further in the following sections.  Finally, if \cx\ is a product of the recent star formation seen in the nearby OB associations, it is unlikely that it is in a system with a low-mass companion, since the secondary would not have had time to evolve off the main sequence and fill its Roche lobe.  We conclude that \cx\ most likely has an intermediate- or high-mass donor if it is related to the nearby star-forming regions.

\subsection{Interpreting the soft excess}\label{specchar}

The spectra of \cx\ are reasonably fit, in most cases, by a power-law model, although in some cases the addition of a soft component (such as a blackbody) is favored with high ($\gtrsim 3\sigma$) significance.  Assuming that this component is due to an accretion disk, we consider the physical implications of the fitted disk blackbody, which has an inner disk temperature of order 0.3 keV.  The observed inner disk temperature in the 2001 October 19 ``bright'' data (0.29 keV) and disk blackbody luminosity (${\rm 1.0\times10^{39}~erg~s^{-1}}$) imply an inner disk radius of $\sim1.2\times10^{3}$ km, which, if taken to be the innermost stable circular orbit of a non-spinning BH, requires a BH mass of $\sim 140~M_{\sun}$ \citep[assuming a Schwarzschild BH.]{makishima00}  Low-temperature disk blackbody components have also been seen in other ULXs, strengthening the hypothesis that they contain IMBHs \citep{miller03}.  

Alternatively, if \cx\ is a stellar-mass black hole (which obeys the
Eddington limit) then we might interpret its high luminosities as due
to some sort of anisotropic emission process.  As discussed in
Section~\ref{location}, the location of the \cx\ near OB 121 is
consistent with an ejected black hole X-ray binary with a primary mass
of 10 $M_{\sun}$.  In this case, the $T_{{\rm in}}$ of $\sim 0.3$ keV
can be compared to the $\sim 1$ keV value found in Galactic BHXN when
they are accreting at their Eddington rate \citep{makishima00}.  If
the disk extends down to the last stable orbit, this lower temperature
implies a lower mass accretion rate, $T_{{\rm in}} \sim  \dot M^{1/4}$
\citep[equation 5.41]{frank92}, which implies that $\dot M$ is
approximately 100 times below Eddington.  Given that the observed peak
disk blackbody luminosity is approximately the Eddington luminosity,
this implies that \cx\ is beamed by a factor of $b \sim 0.01$, where
$b = \Omega/4\pi$.  This is more than one would expect from simple 
scattering of the flux off the inner edge of a funnel shaped accretion disk \citep{king01a}.  

It is therefore difficult to reconcile the high luminosity of \cx\ and
the low temperature of the thermal component without concluding that \cx\ is an intermediate-mass black hole.  However, the presence of a thermal component is not necessarily due to an
accretion disk.   Some portion of the soft flux may be due to another
process, such as  X-rays scattered in an optically-thick outflow \citep{mukai03}.  Or, it may be due to hot diffuse gas associated with the source.  This interpretation is supported by the fact that the soft component is also fit well by a MEKAL model (see, e.g., Roberts et al. 2004 for a discussion of the physical interpretation of MEKAL components in ULX spectra).

\subsection{Spectral and temporal variability and the jet model}

There is some evidence  that the 2001 October 19 spectrum of \cx\ hardens
as it becomes brighter (see the discussion in Section~\ref{spec_fit}).
Note that similar behavior has been observed in other ULXs: for
example, a number of ULXs in NGC 4038/9 (the Antennae) harden with
increasing flux \citep{fabbiano03}, as does NGC 6946 X-11 \citep[MF
16;][]{roberts03b}.  \citet{fabbiano03} posit that this is due to an
increase in the temperature of the inner disk at very high accretion
rates.    However, this model is not consistent with our observations of \cx, since its spectrum is consistently dominated by the power-law component and since
 the variability occurs on extremely short timescales.
\citet{roberts03b} suggest two possible models for the hardening seen
in NGC 6946 X-11---that it may be due to variable emission from the
base of a jet \citep{georganopoulos02}, or that it may be due to
magnetic reconnection events in the accretion disk corona
\citep{dimatteo98}.  We favor the jet model for \cx, since it also
allows for the possibility of relativistic boosting of the X-rays and
thus an amplification of the intrinsic variability.  Relativistic
beaming is a widely accepted explanation for many of the properties of
face-on AGN, and several Galactic jet sources have also demonstrated
variability that may be consistent with beamed X-ray emission:  XTE
J1650$-$500 \citep{tomsick03} and SAX J1819.3$-$2525 \citep[V4641
Sgr,][]{wijnands00}.  

\cx\ exhibits flares which appear relatively equally spaced, and an FFT of the October observation shows that there is a significant broad peak in the low-frequency range of the power spectrum.  However, there is no coherent period, as can also be seen by the multiple peaks in Figure~\ref{period_fold}.  We therefore rule out periodic sources for the variability, such as an orbital or spin modulation.  We consider the possibility of eclipses of the inner disk by clumpy material to be unlikely, since there is no indication of an increase in $N_{\rm H}$ during the low--count-rate periods (it is consistent with the Galactic value throughout).  It is also unlikely that the variability is due to jet precession, since it should then be strictly periodic, which it is not; furthermore, the timescales are very short compared with, for example, the 164 day precession period seen in SS 433 \citep{margon84}.  However, aperiodic variability of this kind could be related to processes at the base of the jet, namely the ejection of relativistic plasma into the jet.  While such processes are not well understood, there is an observed correlation between activity in the inner disk and at the base of the jet in the Galactic microquasar \grs, which happens on timescales similar to what is observed in \cx\ (see Section~\ref{comparisons} for further discussion).

In each of its three observations, \cx\ is extremely variable and exhibits powerful X-ray flares---its luminosity fluctuates by an order of magnitude on timescales of half an hour.  This is suggestive of beaming due to a relativistic jet: as seen in BL Lac objects (commonly called blazars), relativistic beaming along the line of sight leads to rapid, high-amplitude variability.  \cx\ may, in fact, be a ``microblazar''---a microquasar whose relativistic jet is beaming the X-ray emission towards us along the line of sight.  

\subsection{Comparison with other ULXs and Galactic microquasars}\label{comparisons}

Of Galactic sources, \cx\ closely resembles microquasars in both its temporal variability and energetics.  A link between microquasars and ULXs has been posited by a number of authors \citep[for example]{kording02,georganopoulos02}; and recently, radio emission was discovered from a ULX in NGC 5408 \citep{kaaret03}.  This radio emission is almost certainly due to a relativistic jet, and given the high radio flux, it is likely beamed, with a Doppler factor $\delta = \gamma^{-1}(1-\beta\cos\theta)^{-1} \gtrsim 5.8$.  The authors point out that this is consistent with the known ejecta velocities of \grs, which would produce such a Doppler factor given a jet axis tilted up to $10^{\circ}$ with respect to the line of sight (they further note that approximately 1 in 70 extragalactic microquasars should be thus aligned).  The X-ray spectrum of the NGC 5408 source is also consistent with beamed emission, as it is well fit with a broken power-law model (but is also reasonably fit by a disk blackbody of 0.1 keV plus a power-law).

\grs\ is very similar to \cx\ in its temporal characteristics as well as its persistence.   It has been active since its discovery in 1992, exhibiting a wide range of behavior \citep[see, e.g.,][]{muno99, markwardt99}.   In some observations, \grs\ has been observed to vary from $5\times10^{38}$ to ${\rm 3.5\times10^{39}~erg~s^{-1}}$ \citep[assuming $D = 12.5$ kpc]{markwardt99}.  This can be compared to \cx, which has been active since 1980, and varies from $\sim5\times10^{38}$ to ${\rm 1.2\times10^{40}~erg~s^{-1}}$.  Although the low disk temperature and higher peak luminosities observed in \cx\ favor a larger black hole mass, the flares seen in \cx\ occur on the same timescales as some flaring behavior seen in \grs, suggesting a similar dynamical timescale and thus similar physical scale.

Although much of the 0.3--10.0 keV X-ray emission from \grs\ is thought to come from its disk, there is also evidence for a correlation between disk emission and jet formation \citep{mirabel98,eikenberry98}.  Therefore, the timescale for the variability at the base of the jet is likely similar to what is seen in X-ray observations.  The jets of \grs\ are not directed toward our line of sight, but are inclined by 70\arcdeg\ \citep{mirabel94}.  If a jet were directed toward us, we would expect to see relativistically boosted emission varying on similar timescales to what is currently observed: we propose this scenario as a possible explanation for the high-amplitude flares seen in \cx. 

Another prominent feature of Galactic microquasars is their activity at radio wavelengths, which includes variability on timescales of tens of minutes to several hours.  The radio flux of \grs\ has been observed to be as high as 430 mJy at 3.6 cm \citep{rodriguez97}, but at the distance of M74 this would only be 0.8 $\mu$Jy---unfortunately, not detectable by any current radio telescopes.  Even if \cx\ were as radio-bright as the source in NGC 5408, at the distance of M74, it would be at the limit of detectability.  As mentioned above, a search for radio emission in a 1.425 GHz VLA observation did not find emission at the location of \cx\ with an upper limit of 80 $\mu$Jy.

\section{Summary and Conclusions}

\cx\ is a highly variable X-ray source whose 0.3--10.0 keV luminosity is seen to vary from approximately $5\times10^{38}$ to $\rm 1.2\times10^{40}~erg~s^{-1}$ in several thousand seconds.  It was detected by {\it Einstein} in 1980, and observed twice with {\it Chandra} and once with {\it XMM-Newton}.  The latter three observations of \cx\ were each separated by approximately four months.  There have also been a number of optical observations and a radio observation of the region, but as of yet, no counterpart has been found.  Since \cx\ has no optical counterpart and is extremely variable, it is most likely in M74 (and thus a ULX), and may be related to the nearby OB association.  In addition, its proximity to an OB association (a relatively common occurrence for ULXs) implies that it is likely a high- or intermediate-mass system. 

The spectrum of \cx\ is consistent with a low temperature ($\sim 0.3$
keV) disk blackbody component which has been observed in other ULXs.
This component is often interpreted as arising from an accretion disk;
its low temperature then implies an inner disk radius of $1.2 \times
10^{3}$ km and therefore a black hole primary of
mass $140~M_{\sun}$.  However, in many ways \cx\ resembles
stellar-mass Galactic microquasars.  In this scenario, the very high
X-ray luminosity and high-amplitude rapid variability would be
produced by beaming from a jet which is aligned close to our line of
sight.  If \cx\ is, in fact, a stellar-mass black hole of mass $\sim
10~M_{\sun}$, it becomes difficult to explain the low temperature
component in terms of an accretion disk, and it may be from
a different physical source.

\cx\ is an intriguing object.  It appears to be a black hole intermediate- or high-mass X-ray binary where the mass of the black hole remains an open question.  We favor an interpretation in which \cx\ harbors a stellar mass black hole and the high observed X-ray luminosity is produced by beaming.  However, we cannot rule out an intermediate-mass black hole interpretation with the current data.

\section{Acknowledgments}

We would like to thank the anonymous referees for many helpful comments and suggestions.  We thank Jeff McClintock, Phil Kaaret, Andreas Zezas, Andrew King, Joe Patterson, Saul Rappaport, Albert Kong, Peter Freeman, Sebastian Heinz, Sera Markoff, Edo Berger and Chris Stockdale for helpful comments and discussion.  MRG thanks the Aspen Center for Physics for its hospitality.  We also thank the CXC SDS and DS teams, the {\it XMM-Newton} team, and the Gemini Observatory/GMOS team.  Part of this work was based on observations obtained with {\it XMM-Newton}, an ESA science mission with instruments and contributions directly funded by ESA Member States and the USA (NASA). This work was partially supported by NASA contract NAS 8-39073 (CXC) and grant GO1-2092A.


\begin{thebibliography}
\footnotesize

\bibitem[Begelman(2002)]{begelman02} Begelman, M. 2002 ~\apj, 568, L97

\bibitem[Boller~et~al.(1997)]{boller97} Boller, Th., Brandt, W.~N., Fabian, A.~C., \&~Fink, H.~H. 1997, \mnras, 289, 393

\bibitem[Brandt~et~al.(1999)]{brandt99} Brandt, W.~N., Boller, Th., Fabian, A.~C., \&~Ruszkowski, M. 1999, \mnras, 303, L53

\bibitem[Colbert \& Mushotzky(1999)]{colbert99} Colbert, E.~J.~M., \& Mushotzky, R.~F. 1999, \apj, 519, 89

\bibitem[Colbert \&~Ptak(2002)]{colbert02} Colbert, E.~J.~M., \&~Ptak, A.~F. 2002, \apjs, 143, 25

\bibitem[de~Jager \&~Nieuwenhuijzen(1987)]{dejager87} de~Jager, C. \&~Nieuwenhuijzen, H. 1987, \aap, 177, 217

\bibitem[Dewangan~et~al.(2002)]{dewangan02} Dewangan, G.~C., Boller, Th., Singh, K.~P., \&~Leighly, K.~M. 2002, \aap, 390, 65

\bibitem[di~Matteo(1998)]{dimatteo98} di~Matteo, T. 1998, \mnras, 299, L15

\bibitem[Eikenberry~et~al.(1998)]{eikenberry98} Eikenberry, S.~S., Matthews, K., Morgan, E.~H., Remillard, R.~A., \&~Nelson, R.~W. 1998, \apjl, 494, L61~

\bibitem[Esin, McClintock \&~Narayan(1997)]{esin97} Esin, A., McClintock, J., \&~Narayan, R. 1997, \apj, 489, 865

\bibitem[Fabbiano(1989)]{fabbiano89} Fabbiano, G. 1989, \araa, 27, 87

\bibitem[Fabbiano~et~al.(2003)]{fabbiano03} Fabbiano, G., Zezas, A., King, A.~R., Ponman, T.~J., Rots, A. \&~Schweizer, F. 2003, \apjl, 584, L5

\bibitem[Foschini~et~al.(2003)]{foschini03} Foschini, L., Malaguti, G., Di~Cocco, G., Cappi, M., Dadina, M. \&~Ho, L.~C. 2003, in the Proceedings of the 15th SIGRAV Conference on General Relativity and Gravitation (astro-ph/0302143)

\bibitem[Frank, King \&~Raine(1992)]{frank92} Frank, J., King, A., \&~Raine, D. 1992 Accretion Power in Astrophysics, (2d ed., Cambridge: Cambridge Univ. Press)

\bibitem[Georganopoulos, Aharonian~\&~Kirk(2002)]{georganopoulos02} Georganopoulos, M., Aharonian, F.~A. \&~Kirk, J.~G. 2002~A\&A, 388, L25

\bibitem[Giacconi~et~al.(2001)]{giacconi01} Giacconi, R.~et~al. 2001, \apj, 551, 624

\bibitem[Goad~et~al.(2002)]{goad02} Goad, M.~R., Roberts, T.~P., Knigge, C., \&~Lira, P. 2002, \mnras, 335, L67

\bibitem[Huchra, Vogeley \&~Geller(1999)]{huchra99} Huchra, J.~P., Vogeley, M.~S., \&~Geller, M.~J. 1999, \apjs, 121, 287

\bibitem[Irwin, Bregman, \& Athey(2004)]{irwin04} Irwin,~J.~A., Bregman,~J.~N., \& Athey,~A.~E. 2004, \apjl, 601, L143

\bibitem[Ivanov~et~al.(1992)]{ivanov92} Ivanov, G.~R., Popravko, G.,
Efremov, Yu.~N., Tichonov, N.~A., \&~Karachentsev, I.~D.~\ 1992, Astronomy
\&~Astrophysics~Supplement~Series,
96, 645~

\bibitem[Kaaret~et~al.(2003)]{kaaret03} Kaaret, P., Corbel, S., Prestwich, A.~H. \&~Zezas, A. 2003, Science, 299, 365

\bibitem[King~et~al.(2001)]{king01a} King, A.~R., Davies, M.~B., Ward, M.~J., Fabbiano, G., \&~Elvis, M. 2001a, \apjl, 552, L109

\bibitem[K\"{o}rding, Falcke \&~Markoff(2002)]{kording02} K\"{o}rding, E., Falcke, H. \&~Markoff, S. 2002~A\&A, 382, L13

\bibitem[Maccacaro~et~al.(1982)]{maccacaro82} Maccacaro et al. 1982, \apj, 253, 504

\bibitem[Madau \&~Rees(2001)]{madau01} Madau, P. \&~Rees, M.~J. 2001, \apjl, 551, L27

\bibitem[Makishima~et~al.(2000)]{makishima00} Makishima, K.~et~al. 2000, \apj, 535, 632

\bibitem[Margon(1984)]{margon84} Margon, B. 1984, ARA\&A, 22, 507~

\bibitem[Markwardt, Swank \&~Taam(1999)]{markwardt99} Markwardt, C.~B., Swank,
J.~H., \&~Taam, R.~E. 1999, \apjl, 513, L37

\bibitem[Miller~et~al.(2003)]{miller03} Miller, J.~M., Fabbiano, G., Miller, M.~C. \&~Fabian, A.~C. 2003~\apj, 585, L37

\bibitem[Mirabel \&~ Rodr\'\i guez(1994)]{mirabel94} Mirabel, I.~F. \&~ Rodr\'\i guez, L.~F. 1994~Nat., 371, 46

\bibitem[Mirabel~et~al.(1998)]{mirabel98} Mirabel, I.~F., Dhawan, V., Chaty, S., Rodr\'\i guez, L.~F., Marti, J., Robinson, C.~R., Swank, J., Geballe, T. 1998, \aap, 330, L9

\bibitem[Mirabel~et~al.(2002)]{mirabel02} Mirabel, I.~F., Mignani, R., Rodrigues, I., Combi, J.~A., Rodr\'iguez, L.~F. \&~Guglielmetti, F. 2002, A\&A, ~395, 595

\bibitem[Morgan, Remillard \&~Greiner(1997)]{morgan97} Morgan, E.~H., Remillard, R.~A.,
\&~Greiner, J. 1997, \apj, 482, 993

\bibitem[Mukai~et~al.(2003)]{mukai03} Mukai, K., Pence, W.~D., Snowden, S.~L. \&~Kuntz, K.~D. 2003, \apj, 582, 184

\bibitem[Muno, Morgan \&~Remillard(1999)]{muno99} Muno, M.~P., Morgan, E.~H., \&
Remillard, R.~A. 1999, \apj, 527, 321

\bibitem[Pakull \&~Mirioni(2002)]{pakull02} Pakull, M. \&~Mirioni, L. 2002, in Jansen, F. et al., eds, New Visions of the X-ray Universe in the {\it XMM-Newton} and {\it Chandra} Era (astro-ph/0202488)

\bibitem[Podsiadlowski, Rappaport \&~Han(2003)]{podsiadlowski03} Podsiadlowski, Ph., Rappaport, S. \&~Han, Z. 2003, \mnras, 341, 385

\bibitem[Prestwich(2001)]{prestwich01} Prestwich, A. 2001, in~Schlegel, E.~M \&~Vrtilek, S., eds, High~Energy~Universe~at~Sharp~Focus:~{\it~Chandra} Science~(ASP~Conference~Series)

\bibitem[Roberts \&~Warwick(2000)]{roberts00} Roberts, T.~P., \&~Warwick, R.~S. 2000, \mnras, 315, 98

\bibitem[Roberts~et~al.(2002)]{roberts02a} Roberts, T.~P., Goad, M.~R., Ward, M.~J., Warwick, R.~S., \&~Lira, P. 2002a, in Jansen, F. et al., eds, New Visions of the X-ray Universe in the {\it XMM-Newton} and {\it Chandra} Era (astro-ph/0202017)

\bibitem[Roberts~et~al.(2003)]{roberts03} Roberts, T.~P., Goad, M.~R., Ward, M.~J. \&~Warwick, R.~S. 2003, \mnras, 342, 709

\bibitem[Roberts \&~Colbert(2003)]{roberts03b} Roberts, T.~P. \&~Colbert, E.~J.~M 2003, \mnras, 341, L49

\bibitem[Roberts et al.(2004)]{roberts04} Roberts, T.~P., Warwick, R.~S., Ward, M.~J., \& Goad, M.~R. 2004, \mnras, 349, 1193
 
\bibitem[Rodr\'\i guez \&~Mirabel(1997)]{rodriguez97} Rodr\'\i guez, L.~F. \&~Mirabel, I.~F. 1997, \apjl, 474, L123

\bibitem[Stockdale(2002)]{stockdale02} Stockdale, C. 2002, private~communication

\bibitem[Stocke~et~al.(1991)]{stocke91} Stocke, J.~T., Morris, S.~L., Gioia, I.~M., Maccacaro, T., Schild, R., Wolter, A., Fleming, T.~A., \&~Henry, J.~P. 1991, \apjs, 76, 813

\bibitem[Tomsick~et~al.(2003)]{tomsick03} Tomsick, J.~A., Kalemci, E., Corbel, S. \&~Kaaret, P. 2003, \apj, 592, 1100

\bibitem[van~Paradijs \&~McClintock(1994)]{vanparadijs94} van~Paradijs, J. \&~McClintock, J.~E.~\ 1994, \aap, 290, 133

\bibitem[Wijnands \&~van~der~Klis(2000)]{wijnands00} Wijnands, R. \&~van~der~Klis, M. 2000, \apjl, 528, L93

\bibitem[Zezas \&~Fabbiano(2002)]{zezas02} Zezas, A. \&~Fabbiano, G. 2002, \apj, 577, 726

\end{thebibliography}
\end{document}